\documentclass[preprint,flushrt]{aastex}

\newcommand{\caii}{\ion{Ca}{2}}
\newcommand{\civ}{\ion{C}{4}}
\newcommand{\hi}{\ion{H}{1}}
\newcommand{\hinv}{\ensuremath{h^{-1}_{70}}}
\newcommand{\kms}{{\rm km~s}\ensuremath{^{-1}}}
\newcommand{\Lstar}{\ensuremath{\,{\rm L}^*}}
\newcommand{\lya}{{\rm Ly}\ensuremath{\alpha}}
\newcommand{\nai}{\ion{Na}{1}}
\newcommand{\ngc}{NGC~3067}
\newcommand{\qso}{3C~232}
\newcommand{\siiv}{\ion{Si}{4}}
\newcommand{\Zsun}{\ensuremath{\,Z_{\Sun}}}

\begin{document}

\title{Absorption Line Study of Halo Gas in \ngc\ Toward the Background Quasar \qso}
\author{Brian A. Keeney\altaffilmark{1}, Emmanuel Momjian\altaffilmark{2}, John T. Stocke\altaffilmark{1}, Chris L. Carilli\altaffilmark{3}, \\and Jason Tumlinson\altaffilmark{4}}
\altaffiltext{1}{Center for Astrophysics and Space Astronomy, Department of Astrophysical and Planetary Sciences, Box 389, University of Colorado, Boulder, CO 80309}
\altaffiltext{2}{National Astronomy and Ionosphere Center, Arecibo Observatory, HC3 Box 53995, Arecibo, PR 00612}
\altaffiltext{3}{National Radio Astronomy Observatory, P.O. Box O, Socorro, NM 87801}
\altaffiltext{4}{Department of Astronomy and Astrophysics, University of Chicago, Chicago, IL 60637}
\email{keeney@colorado.edu, emomjian@naic.edu, stocke@casa.colorado.edu, ccarilli@nrao.edu, tumlinso@oddjob.uchicago.edu}

\shorttitle{Absorption in \ngc}
\shortauthors{Keeney et al.}

\begin{abstract}

We present new \hi\ 21~cm absorption data and ultraviolet spectroscopy from HST/STIS of the QSO/galaxy pair \qso/\ngc. The QSO sightline lies near the minor axis and $1\farcm8$ (11\hinv~kpc) above the plane of \ngc, a nearby luminous ($cz = 1465 \pm 5$~\kms, ${\rm L} = 0.5$\Lstar) starburst galaxy with a moderate star formation rate of 1.4~M$_{\Sun}$~yr$^{-1}$. The UV spectra show that the \siiv\ and \civ\ doublets have the same three velocity components at $cz = 1369$, 1417, and 1530~\kms\ found in \caii\ H \& K, \nai\ D, \ion{Mg}{1}, \ion{Mg}{2}, and \ion{Fe}{2}, implying that the low and high ionization gas are both found in three distinct absorbing clouds (only the strongest component at 1420~\kms\ is detected in \hi\ 21~cm). The new \lya\ observation allow the first measurements of the spin and kinetic temperatures of halo gas: $T_{\rm s} = 435 \pm 140$~K and $T_{\rm k}/T_{\rm s} \approx 1$. However, while a standard photoionization model can explain the low ions, the \civ\ and \siiv\ are explained more easily as collisionally-ionized boundary layers of the photoionized clouds. Due to their small inferred space velocity offsets ($\Delta v = -260$, $-130$, and +170~\kms) relative to the nucleus of \ngc\ and the spatial coincidence of low and high ionization gas, we propose that these absorbers are analogous to Galactic high velocity clouds (HVCs). A comparison of the \ngc\ clouds and Galactic HVCs finds similar \hi\ column densities, kinematics, metallicities, spin temperatures, and inferred sizes. We find no compelling evidence that any halo gas along this sightline is escaping the gravitational potential of \ngc, despite its modest starburst.

\end{abstract}


\section{Introduction}
\label{intro}

Starburst-driven superwinds, powered by the cumulative effect of coeval supernova explosions, have long been suggested as a mechanism by which metals and energy can be transported from galaxies to enrich their surroundings \citep[e.g.,][]{deyoung78,heckman98}. However, observations have yet to prove that starburst galaxies commonly generate winds that are capable of escaping the galaxy's gravitational potential. Most studies measure the wind velocity near the disk of a starburst galaxy. Using such velocities to try to determine whether the wind is gravitationally bound to the galaxy is not straightforward, however, since the expanding superbubble will inevitably sweep up gas as it travels through the galactic halo, causing the wind to decelerate.

The spatial extent of wind gas can be measured by H$\alpha$ or X-ray imaging \citep{watson84,martin99,martin02} and follow-up spectroscopy can determine the wind velocity. The X-ray images themselves allow one to measure the wind temperature, which is found to be nearly constant as a function of a galaxy's maximum \hi\ rotation speed \citep{martin99}, suggesting that superwinds are more likely to escape from dwarf galaxies than more massive ones. These emission line studies are limited to studying the densest regions of wind gas, however, and cannot measure the full extent of the gas or the wind velocity in the diffuse halo. 

Given a suitably bright background source, typically the stellar continuum of the starburst region itself, high velocity wind gas can also be detected in absorption \citep[e.g.,][]{heckman00,heckman01}. Such studies detect typical outflow velocities of 400-1000~\kms, although the outflow velocities in dwarf galaxies can be considerably lower \citep{heckman00,martin03}. While absorption line studies are more sensitive to diffuse gas than emission line studies, they suffer from an ambiguity in the distance between the background source and the absorbing gas. These studies can only definitively state that the absorber lies in front of the background source and yield no information about how far in front of the source the absorber lies. Since absorption line studies of galactic winds typically use the starburst itself as a background source, they cannot differentiate between a high velocity outflow in the galactic halo or within the starburst region.

Currently, the only method that directly measures the properties of the superwind gas well away from its parent galaxy relies on searching for absorption from the wind gas against a bright background QSO \citep[e.g.,][]{stocke04}. Unfortunately, coincidences of a bright (${\rm m_V} \le 17$) QSO and a low-$z$ galaxy separated by $\la 150$\hinv~kpc in projection are exceedingly rare; only $\sim 60$ are known to exist over the entire sky, and most of the galaxies in these pairs are not starbursts. Thus, a systematic study of galaxy winds with this technique is not yet possible, although much can be learned by observing those QSO/galaxy pairs that do exist.

Rest-frame ultraviolet absorption line spectroscopy of bright QSOs is a very sensitive tracer of diffuse gas in interstellar and intergalactic space. For example, the weakest \lya\ absorption lines detected by the {\it Hubble Space Telescope} (HST) probe gas at \hi\ column densities, $N_{\rm HI} \sim 10^{13}~{\rm cm}^{-2}$, comparable to the weakest \lya\ lines detected in Keck and VLT ground-based spectra at high redshift. Associated metal-line absorptions are commonly found in individual \hi\ absorbers down to $N_{\rm HI} \sim 10^{14}~{\rm cm}^{-2}$ and pixel addition techniques cumulatively detect \ion{Si}{3}, \civ, and \ion{O}{6} absorptions down to $N_{\rm HI} \sim 10^{13}~{\rm cm}^{-2}$ \citep{schaye03,aguirre04,simcoe04}. Given that the number density of \lya\ absorbers greatly increases both with increasing redshift and with decreasing column density, a ``unified'' model for these absorbers places them all in the extended halos of bright galaxies \citep[e.g.,][]{steidel95,lanzetta95}. However, the radii of such halos become implausibly large at $z \sim 2$ for the weak ($N_{\rm HI} \sim 10^{14}~{\rm cm}^{-2}$) \lya\ and metal line systems: $\ge 150$\hinv~kpc for L$^*$ galaxies. A more recent interpretation \citep*{stocke05} suggests that, while the higher column density absorbers could be the bound halos of bright galaxies, the weaker absorbers arise in unbound galaxy winds produced by the much more numerous low luminosity galaxies (${\rm M_B} \ge -16$).

QSO absorption line systems with column densities in the range $N_{\rm HI} = 10^{17.3}$ to $10^{20.3}$~cm$^{-2}$, known as Lyman limit systems (LLSs), are associated with bright (${\rm L} > 0.1$-0.3\Lstar) galaxy halos as shown by \citet{steidel95,steidel98}. \citeauthor{steidel95} found that 55 of the 58 LLS absorbers with \ion{Mg}{2} equivalent widths $\ge 0.3$~\AA\ in his sample at $0.2 \le z \le 1$ were associated with bright galaxies at the same redshift within a projected distance of 50\hinv~kpc. Follow-up spectra of the underlying stellar kinematics in these galaxies \citep{steidel02} and HST imaging and UV spectra of the \civ\ absorption in these systems \citep{churchill00} further indicate that LLSs are the bound gaseous halos of luminous galaxies and that in many cases these halos share the kinematics of the underlying stellar disk.

\qso/\ngc\ is a well-studied QSO/galaxy pair in which the QSO sightline probes the gaseous galaxy halo of a low-$z$ spiral galaxy \citep{burbidge71}. The \qso\ sightline lies near the minor axis of \ngc\ and is located $1\farcm8$ (11\hinv~kpc projected) above the galactic plane. \ngc\ is a luminous \citep*[${\rm M_B} \approx -19$, ${\rm L} \approx 0.5$\Lstar;][]{devaucouleurs91,marzke94} nearly edge-on spiral \citep*[$i = 68^{+4}_{-3}$~degrees;][]{rubin82} with a radial velocity of $cz = 1465 \pm 5$~\kms\ (Carilli \& van Gorkom 1992, hereafter \citet{cvg}) and a far infrared luminosity of ${\rm L_{FIR}} = 2.4 \times 10^{43}$~ergs~s$^{-1}$ \citep*{shapley01}, which corresponds to a star formation rate (SFR) of 1.4~M$_{\Sun}$~yr$^{-1}$ \citep{misiriotis04}. The SFR of \ngc\ is somewhat less than that of other nearby luminous (${\rm L} \approx {\rm L}^*$) starbursts like NGC~253 (${\rm SFR} = 3.7~{\rm M_{\Sun}~yr^{-1}}$), NGC~3079 (${\rm SFR} = 6.4~{\rm M_{\Sun}~yr^{-1}}$), and NGC~4945 \citep[${\rm SFR} = 3.1~{\rm M_{\Sun}~yr^{-1}}$;][]{devaucouleurs91,marzke94,shapley01,misiriotis04,grimm03}. Thus, one would expect \ngc\ to produce a wind similar to but perhaps weaker than those of other starburst galaxies in the local Universe.

\hi\ 21~cm absorption against \qso\ was first detected near the redshift of \ngc\ at $cz = 1418 \pm 2$~\kms\ by \citet{haschick75}, who conjectured that the absorption was caused by \hi\ in the outermost regions of \ngc. \citet{cvg} observed \qso/\ngc\ in \hi\ 21~cm emission at $50\arcsec$ resolution using the  VLA and found \hi\ distributed slightly asymmetrically (the 21~cm emission extends further to the east of the kinematic center than to the west, and the kinematic center is offset from the optical center) at velocities ranging from 1310 to 1635~\kms. \citet{cvg} also found a region of extended emission north and south of \ngc\ with velocities ranging from 1350 to 1550~\kms, from which the 21~cm absorption seen toward \qso\ arises. \citet{cvg} speculated that the extended emission is caused either by extended spiral arms, which may be the result of a past encounter with an as-yet undetected companion, or by viewing gas in a polar ring that is seen nearly face-on. In the former case we would expect the \hi\ seen above and below the plane of \ngc\ to be rotating in the same direction as the underlying disk; in the latter case the space velocity field is less well-constrained.

Metal lines have also been detected along the \qso\ sightline at the same velocity as the \hi\ 21~cm absorption found by \citeauthor{haschick75}. \citet{boksenberg78} found \caii\ H and K at $cz = 1406 \pm 11$~\kms\ with the Hale 5m. Stocke et al. (1991, hereafter \citet{s91}) used the Multiple Mirror Telescope to find three separate velocity components in \caii\ H \& K and \nai\ D at $cz = 1369 \pm 2, 1417 \pm 2$, and $1530 \pm 10$~\kms\ and Tumlinson et al. (1999, hereafter \citet{t99}) used the Goddard High Resolution Spectrograph (GHRS) aboard HST to discover that \ion{Mg}{1}, \ion{Mg}{2}, and \ion{Fe}{2} absorption lines show evidence for the same three velocity components that \citet{s91} found in \nai\ and \caii, although the \ion{Mg}{2} lines are highly saturated and could cover this entire velocity range. None of these spectra preclude weaker absorption at velocities intermediate to the three strong systems.

The \ngc\ halo cloud probed by the \qso\ sightline has been used to set both an upper limit based upon H$\alpha$ imaging \citep[${\rm I}_{\nu} \le 4 \times 10^{-23}~{\rm ergs~s^{-1}~cm^{-2}~Hz^{-1}~sr^{-1}}$;][]{s91} and a lower limit based upon \ion{Mg}{2}/\ion{Mg}{1} absorption line ratios \citep[${\rm I}_{\nu} \ge 1 \times 10^{-23}~{\rm ergs~s^{-1}~cm^{-2}~Hz^{-1}~sr^{-1}}$;][]{t99} on the extragalactic ionizing flux (I$_{\nu}$). The H$\alpha$ images also yielded a limit on the escape fraction of ionizing radiation from the disk of \ngc\ to the halo locations probed by the \qso\ sightline of $f_{esc} \le 2$\% \citep{t99}. This value is comparable to the most recently derived $f_{esc} = 1$-2\% for the Galaxy from H$\alpha$ images of high velocity clouds \citep[HVCs;][]{putman03}.

This paper presents \hi\ 21~cm and ultraviolet spectra of \qso\ to determine if the multiple velocity components found by \citet{s91} and \citet{t99} along this sightline are present in \hi\ \lya, \siiv, and \civ. In \S\ref{obs} we describe our radio and UV observations and analysis. In \S\ref{discussion} we hypothesize that the multiple velocity components found by \citet{s91} correspond to HVCs around \ngc\ and compare the properties of these clouds to those of Galactic HVCs. In \S\ref{conclusions} we summarize our findings and discuss the evidence for unbound starburst winds in light of our results.

\section{Observations and Data Reduction}
\label{obs}

Our full dataset includes previously-published optical and UV spectra of \qso\ \citep{s91,t99} and published Very Large Array (VLA) \hi\ emission maps \citep{cvg}, as well as new \hi\ 21~cm observations with the Very Long Baseline Array (VLBA) and the Arecibo radio telescope and new HST low and medium resolution far-UV spectra obtained with the Space Telescope Imaging Spectrograph (STIS). The \hi\ 21~cm and HST/STIS observations are described below, while the previously published data will be described as needed in \S\ref{discussion}.

\subsection{VLBA Observations}
\label{obs:vlba}

High angular resolution observations of \qso\ were carried out at 1413.7~MHz on 2002 January 31 using the VLBA, which is operated by NRAO\footnote[5]{The National Radio Astronomy Observatory (NRAO) is a facility of the National Science Foundation operated by Associated Universities, Inc.}. The bandwidth of the observations was 8~MHz, in each of the right- and left-hand circular polarizations, sampled at two bits and centered at the frequency of the \hi\ 21~cm line, at a heliocentric redshift of $z=0.004736$, or $cz=1420$~\kms. The data were correlated at the VLBA correlator in Socorro, New Mexico, with 1024-point spectral resolution per baseband channel and 4~s correlator integration time. The resulting channel separation was 7.8~kHz (1.7~\kms). The total observing time was 11.7~hr. Along with \qso, the radio source J0927+3902 was observed at the same frequency as a fringe finder and a bandpass calibrator. Data reduction and analysis were performed using the Astronomical Image Processing System (AIPS), and the Astronomical Information Processing System (AIPS++), of NRAO.

\begin{figure}[!t]
\begin{center}
\epsscale{1.0}
\plotone{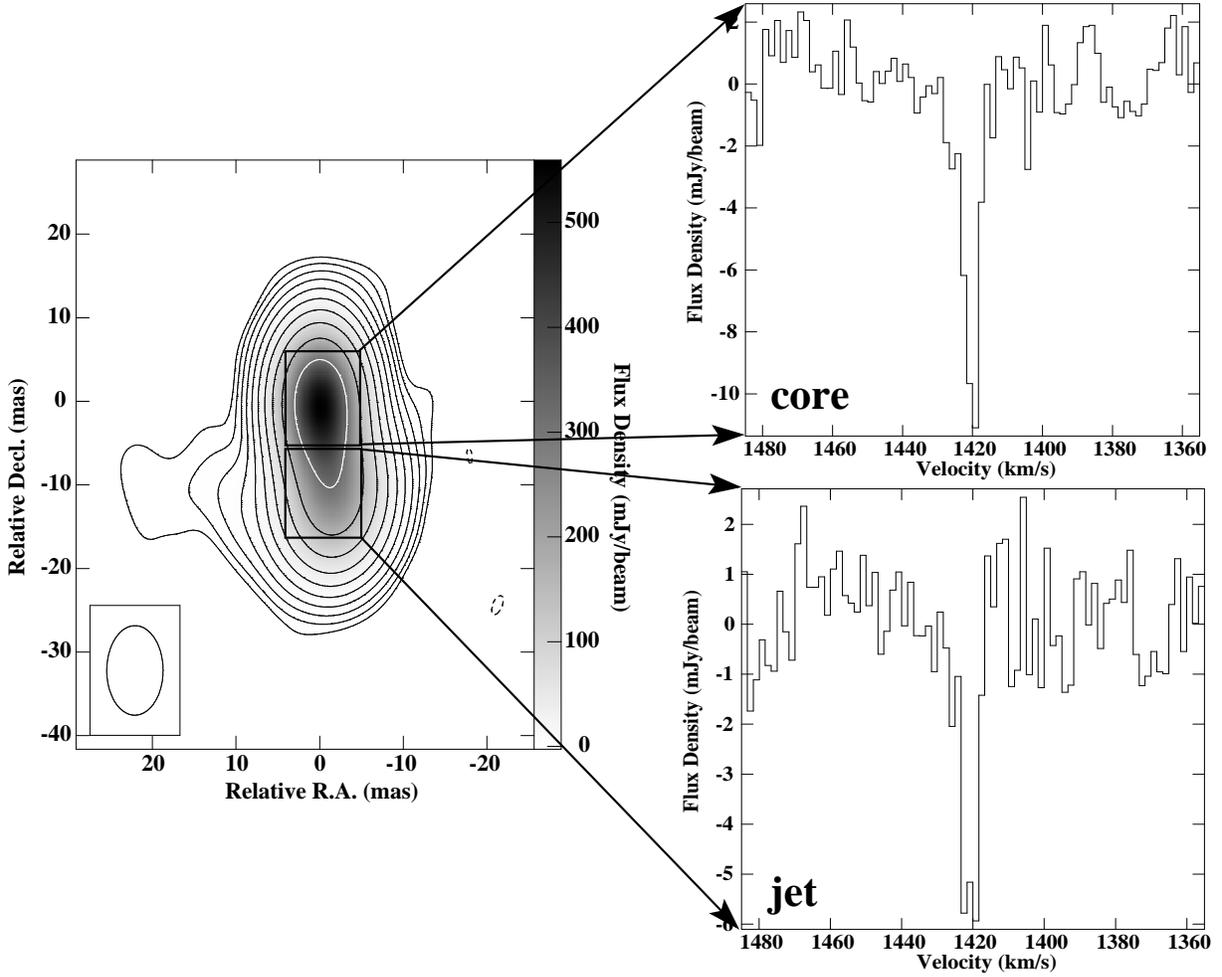}
\end{center}
\caption{\hi\ absorption spectra obtained at two locations against the background continuum source \qso\ at 1413.7~MHz. No absorption $> 3\sigma$ is detected toward other continuum emission regions in \qso. The restoring beam size is $10.7 \times 6.7$~mas in position angle $-0.2\degr$. The rms noise level in the \hi\ spectra is 1.6~mJy~beam$^{-1}$. The velocity resolution is 1.7~\kms. The contour levels of the continuum image are $-5$, 5, 10, 20, 40, ..., 2560 times the rms noise level, which is 123~$\mu$Jy~beam$^{-1}$. The reference position (0,0) is $\alpha$(J2000) = $09^{\rm h}~58^{\rm m}~20\fs9499$, $\delta$(J2000) = $32\degr~24\arcmin~02\farcs177$. At the redshift of \ngc, $1~{\rm mas} = 0.1$~pc for $H_0 = 70$~\kms~Mpc$^{-1}$.
\label{fig:HIspec}}
\end{figure}

\begin{figure}[!t]
\begin{center}
\epsscale{1.0}
\plotone{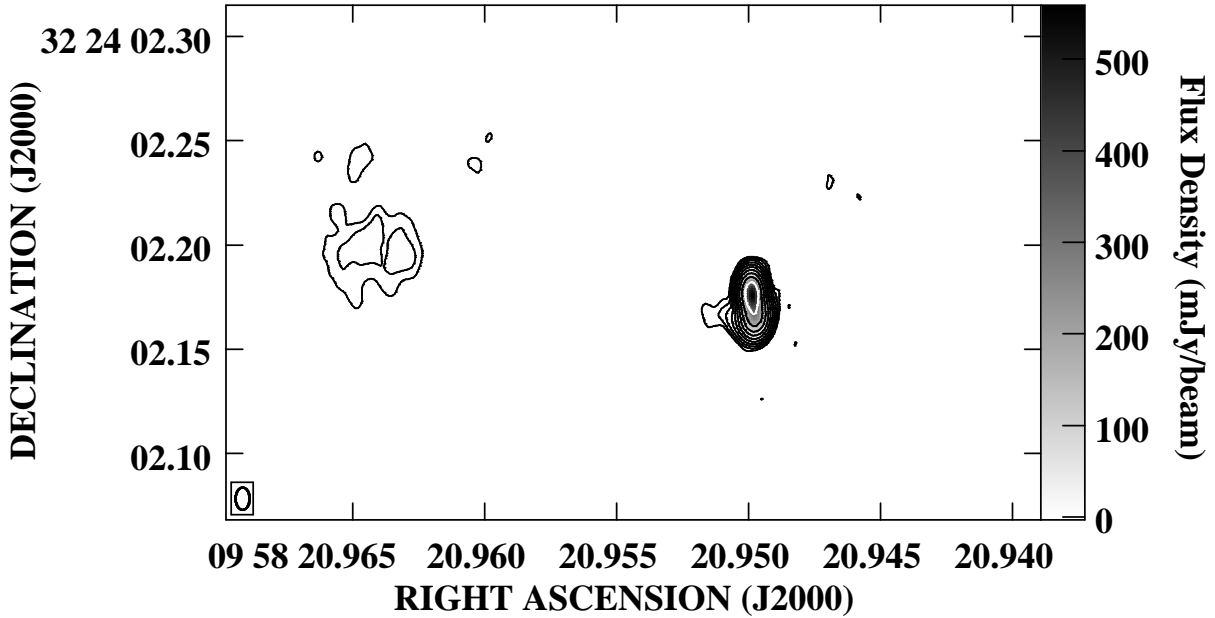}
\end{center}
\caption{A large-field continuum image of \qso\ at 1413.7~MHz. The restoring beam and contour levels are the same as in Figure~\ref{fig:HIspec} and the peak flux is 560~mJy~beam$^{-1}$. Note the diffuse continuum emission $\sim 180$~mas to the east of \qso.
\label{fig:wide}}
\end{figure}

Figure~\ref{fig:HIspec} shows our high resolution L-band continuum image of \qso\ with $10.7 \times 6.7$~mas resolution, along with \hi\ absorption spectra averaged over two regions. The image was obtained by using the CLEAN algorithm as implemented in AIPS, with an intermediate grid weighting (Robust=+2 in IMAGR). The continuum source consists of a compact core and a jet structure extending to the south, with a weaker extension to the east (but see Figure~\ref{fig:wide}). A small amount of additional, very low surface brightness flux exists on the 500~mas scale \citep{condon98}. Table~\ref{tab:vlbafits} lists the results of fitting Gaussian models to the observed spatial profile for the source components listed in column 1. We designate the two regions of \qso\ toward which the spectra in Figure~\ref{fig:HIspec} have been extracted as core and jet. No \hi\ absorption larger than $3\sigma = 4.7$~mJy~beam$^{-1}$ is detected toward other continuum regions in \qso. The velocity full width at half maximum (FWHM) of the \hi\ absorption is $4.7 \pm 0.1$~\kms\ centered at $1420.0 \pm 0.1$~\kms. No \hi\ absorption is detected at 1369 and 1530~\kms, in contrast to where metal lines appear in optical and UV data \citep{s91,t99}.

\begin{deluxetable}{cccccccc}

\tablecolumns{8}
\tablewidth{0pt}

\tablecaption{Gaussian Fits to the Continuum Components in \qso\ (Fig.~\ref{fig:HIspec})
\label{tab:vlbafits}}

\tablehead{\colhead{} &  & \colhead{Relative Position\tablenotemark{a}} & \colhead{Peak} & \colhead{Total} & \colhead{Major\tablenotemark{b}} & \colhead{Minor\tablenotemark{b}} &\colhead{P.A.} \\ \colhead{Source} & \colhead{} & \colhead{(mas)} & \colhead{(mJy~beam$^{-1}$)} & \colhead{(mJy)} & \colhead{(mas)} & \colhead{(mas)}& \colhead{($\degr$)}}

\startdata
core\dotfill & &   0 , 0  & $478.2 \pm 0.3$ & $587.0 \pm 0.3$ & 8.7 & 4.7 & $175 \pm 5$ \\
jet\dotfill & &  1.5W, 11S & $196.2 \pm 0.5$ & $388.2 \pm 0.9$ & 11.0 & 6.1 & $179 \pm 5$ \\
east.~ext.\dotfill & & 7.5E, 14S  & $3.92 \pm 0.02$ & $4.66 \pm 0.03$ & 8.3 & 4.8 & $43 \pm 20$ \\
\enddata

\tablenotetext{a}{\ The position (0,0) is $\alpha\rm{(J2000)} = 09^{\rm h}~58^{\rm m}~20\fs9499$, $\delta\rm{(J2000)} = +32\degr~24\arcmin~02\farcs177$.}
\tablenotetext{b}{\ Errors are $\approx 1$~mas}

\end{deluxetable}

Figure~\ref{fig:wide} is a wide field continuum image with the same angular resolution as Figure~\ref{fig:HIspec}, showing an extended structure at an angular distance of $\sim 180$~mas to the east of the main continuum structure of \qso. A gaussian fit to this structure results in a total flux density of $46.9 \pm 0.4$~mJy, and a size at full width half maximum of $60 \times 36$~mas on the plane of the sky. No \hi\ absorption larger than 3$\sigma = 4.7$~mJy~beam$^{-1}$ is detected toward this diffuse continuum emission region. 

The peak optical depth values in the core and the jet regions are $3.5 \pm 0.8$\% and $2.6 \pm 0.8$\%, respectively, which yields column densities ($N_{\rm HI}$) of \hi\ absorption of $N_{\rm HI}/T_{\rm s} = 1.7 (\pm 0.4) \times 10^{17}~{\rm cm}^{-2}~{\rm K}^{-1}$ and $1.1 (\pm 0.3) \times 10^{17}~{\rm cm}^{-2}~{\rm K}^{-1}$ for the core and jet regions, respectively. An absorber with this column density would cause a flux decrement of $\sim 0.1$~mJy~beam$^{-1}$ against the eastern extension of the jet (see Figure~\ref{fig:HIspec}) and $\sim 1$~mJy~beam$^{-1}$ against the diffuse emission region in Figure~\ref{fig:wide}. Since the $3\sigma$ detection limit in these regions is 4.7~mJy~beam$^{-1}$, we cannot exclude the possibility that the \hi\ absorption is constant over a region extending at least 20~mas (2\hinv~pc) and possibly as large as $0.2\arcsec$ (20\hinv~pc) on the sky.

\citet{cvg} measured the column density of \hi\ 21~cm emission toward \qso\ to be $8(\pm 4) \times 10^{19}~{\rm cm}^{-2}$ using the VLA ($50\arcsec$ beam). Comparing this value with the $N_{\rm HI}/T_{\rm s}$ derived from our VLBA \hi\ absorption observations, we derive a hydrogen spin temperature of $T_{\rm s} = 470 \pm 260~{\rm K}$. Of course, this value assumes that the \hi\ 21~cm emission and absorption have comparable physical extent.

\subsection{Arecibo Observations}
\label{obs:arecibo}

Single dish observations of \qso/\ngc\ were carried out with the L-band Wide receiver of the 305~m Arecibo radio telescope\footnote[6]{The Arecibo Observatory is part of the National Astronomy and Ionosphere Center, which is operated by Cornell University under a cooperative agreement with the National Science Foundation.} on 2003 November 23-24, for a total of 2.5~hr. The double position switching observations utilized two of the interim correlator boards to observe at the frequency of the \hi\ 21~cm line in \ngc. The bandwidth of each board was 6.25~MHz, with one linear polarization each and 2048 spectral channels. The source 4C+32.34 was observed as a bandpass calibrator.

\begin{figure}[!t]
\begin{center}
\epsscale{0.95}
\plotone{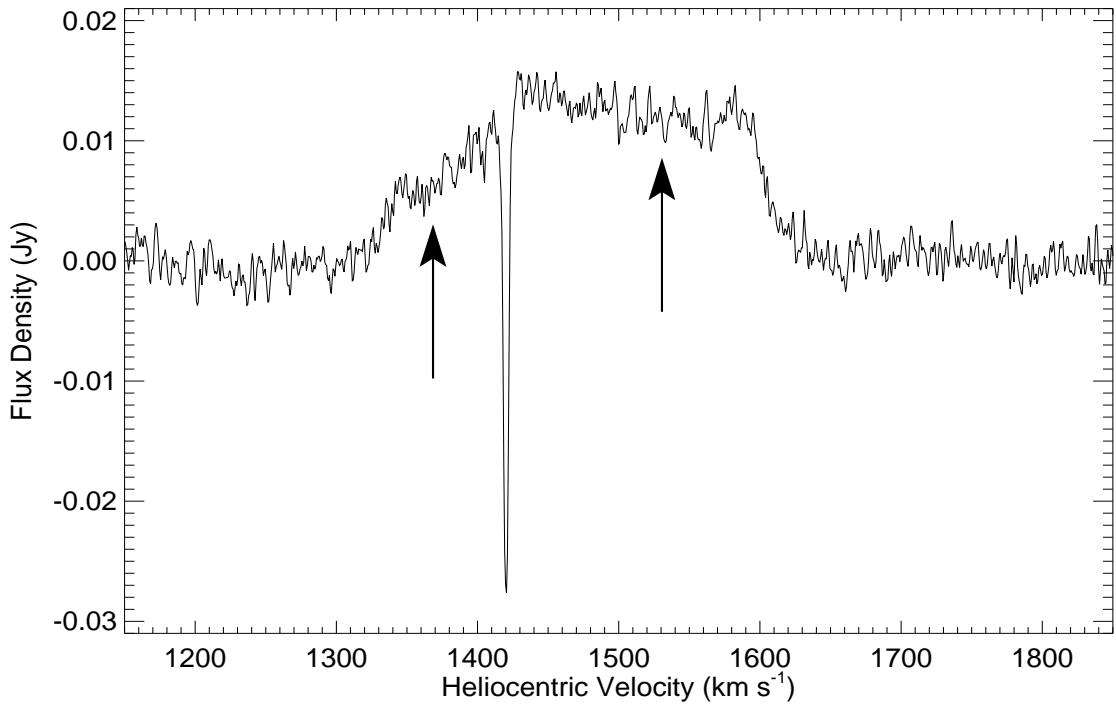}
\end{center}
\caption{Hanning-smoothed \hi\ 21~cm spectrum of the galaxy \ngc\ taken at Arecibo. The velocity resolution is 1.3~\kms, and the rms noise level is 1.2~mJy. The arrows indicate the positions of the other two velocity components ($cz = 1369$, 1530~\kms) found in various metal species \citep{s91,t99}. Note that all three radial velocities fall well within the \hi\ emission profile of \ngc, and that the \hi\ 21~cm absorption detection is close to the galactocentric radial velocity of 1465~\kms\ \citep{cvg}.
\label{fig:arecibo}}
\end{figure}

Figure~\ref{fig:arecibo} is a total intensity Hanning-smoothed \hi\ spectrum of \qso/\ngc. The plot clearly shows the $\approx 300$~\kms\ wide \hi\ 21~cm emission in \ngc,  and the narrow \hi\ 21~cm absorption line that arises on the line of sight to the background quasar \qso\ at $1420.2 \pm 0.1$~\kms. The velocity resolution of this spectrum is 1.3~\kms, and the rms noise level is 1.2~mJy. The velocity FWHM of the 21~cm absorption is $4.2 \pm 0.1$~\kms\ and its peak optical depth is $2.81 \pm 0.08\%$, yielding a column density of $N_{\rm HI}/T_{\rm s} = 2.3(\pm 0.1) \times 10^{17}~{\rm cm}^{-2}~{\rm K}^{-1}$. While the single dish optical depth is marginally larger than the VLBA measurements, these observations suggest no strong variation in covering factor or $T_{\rm s}$ over a size scale of $0\farcs2$, or 20\hinv~pc. Again using the \hi\ 21~cm emission column density found by \citet{cvg}, this value corresponds to a hydrogen spin temperature of $T_{\rm s} = 350 \pm 175$~K, which agrees with the spin temperature calculated from the VLBA column density to within errors.

No \hi\ absorption is detected at the two other metal-line velocities to $\tau < 0.24\%~(3\sigma)$. Assuming a spin temperature of $T_{\rm s} = 450$~K and the same velocity width as the 1420~\kms\ absorption line, this optical depth limit corresponds to a $3\sigma$ column density limit of $N_{\rm HI} < 9 \times 10^{18}$~cm$^{-2}$. Notice that the three metal line velocities are well within the extent of the disk rotation curve.

\begin{deluxetable}{lcccc}

\tablecolumns{5}
\tablewidth{0pt}

\tablecaption{Journal of the HST/STIS Observations Toward \qso
\label{tab:stisobs}}

\tablehead{\colhead{} & \colhead{Wavelength Coverage} & \colhead{Resolution} & \colhead{} & \colhead{Usable Exposure Time} \\ \colhead{Setting} & \colhead{(\AA)} & \colhead{(\kms)} & \colhead{Date of Observation} & \colhead{(s)}}

\startdata
1\dotfill & 1372-1428 &  20 & 2001 Oct 29    & 11,178 \\
2\dotfill & 1194-1250 &  25 & 2001 Oct 30-31 & 11,178 \\
3\dotfill & 1540-1594 &  20 & 2002 Jan 02    &  8,208 \\
          & 1540-1594 &  20 & 2002 Jan 04    & 14,098 \\
4\dotfill & 1150-1700 & 250 & 2003 Mar 04    &  1,200 \\
\enddata

\end{deluxetable}

\subsection{HST/STIS Observations}
\label{obs:stis}

\begin{figure}[!t]
\begin{center}
\epsscale{0.95}
\plotone{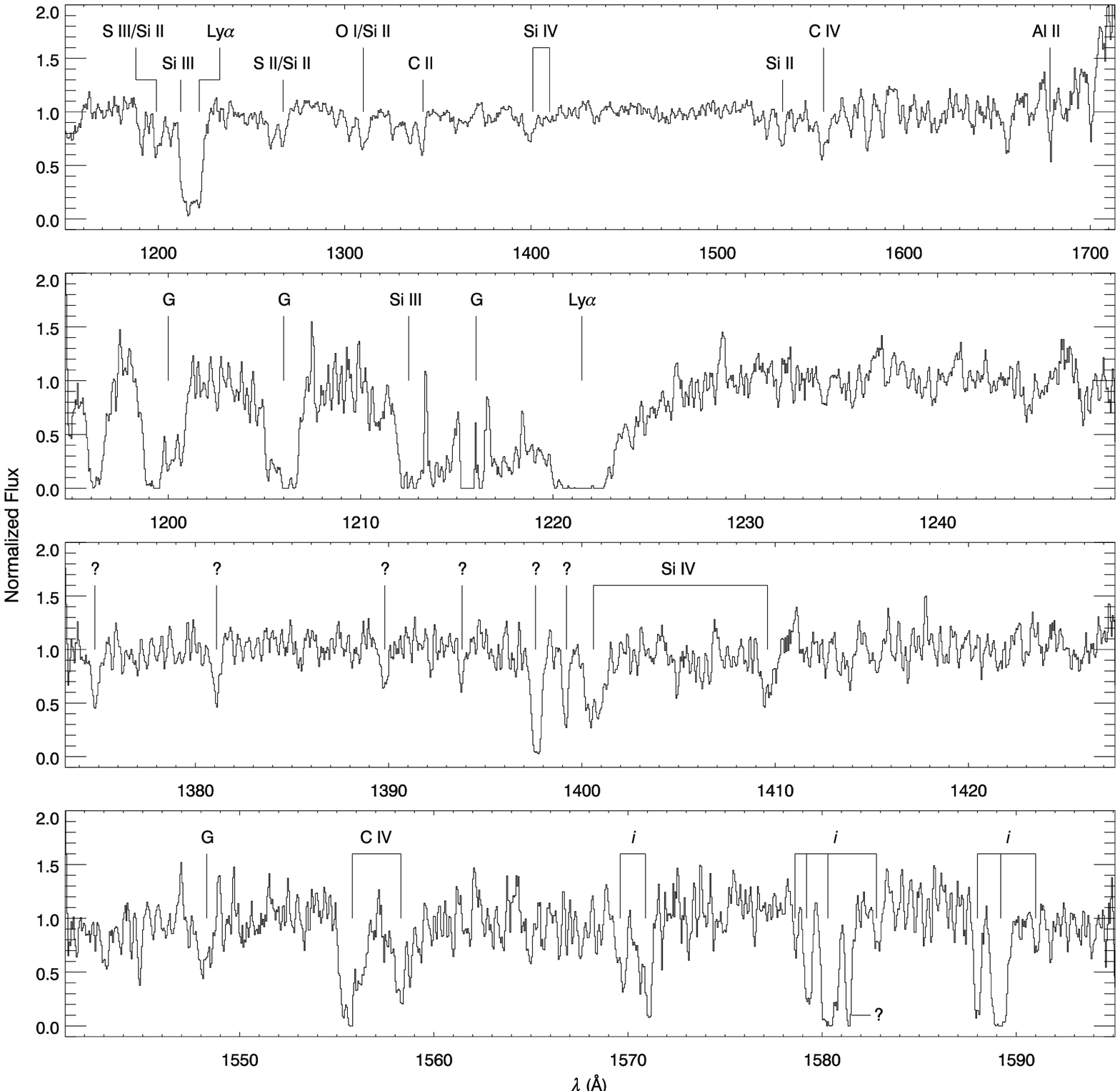}
\end{center}
\caption{The final coadded, normalized STIS spectra. The top panel shows the G140L snapshot spectrum, the second panel shows the G140M spectrum with a central wavelength of 1222~\AA, the third panel shows the G140M spectrum centered at 1400~\AA, and the bottom panel shows the G140M spectrum centered at 1567~\AA. Absorption features from \ngc\ are labelled in each panel; in the G140L spectrum, the \ion{Si}{3} and \lya\ absorption from \ngc\ are blended with Galactic \lya\ absorption. In the G140M spectra, features marked with a ``G'' are absorption from our Galaxy and those marked with an ``{\it i}'' are absorption intrinsic to \qso\ at $z = 0.533$. Features marked with a ``?'' are likely low-$z$ \lya\ absorbers since the \lya\ forest region of \qso\ extends to 1864~\AA.
\label{fig:normspec}}
\end{figure}

\qso\ was observed by HST/STIS\footnote[7]{Based on observations with the NASA/ESA {\it Hubble Space Telescope}, obtained at the Space Telescope Science Institute, which is operated by the Association of Universities for Research in Astronomy, Inc., under NASA contract NAS5-26555.} with the G140L grating as part of GO Snapshot Program 9506 (J. Stocke, PI), and also with the G140M grating at three wavelength settings as part of GO Program 8596 (M. Pettini, PI). A journal of these observations appears in Table~\ref{tab:stisobs}; total exposure times were 1.2~ksec with G140L, 11.2~ksec with G140M at central wavelengths of 1222 and 1400~\AA, and 22.3~ksec with G140M at a central wavelength of 1567~\AA. Multiple exposures with the same central wavelength were shifted to a heliocentric reference frame and coadded with signal-to-noise weighting. The continuum in each wavelength region was normalized using Legendre polynomials and the resulting spectra were examined for evidence of absorption from \ngc. These normalized spectra are shown in Figure~\ref{fig:normspec} and absorption lines associated with \ngc\ are labelled at their observed wavelengths.

The G140L spectrum yields data on \lya, \ion{O}{1}, \ion{C}{2}, \civ, \ion{Si}{2}, \ion{Si}{3}, \siiv, \ion{S}{2}, \ion{S}{3}, and \ion{Al}{2}. All of the prominent Si and C lines in the wavelength range covered by this spectrum are detected over a wide range of ionization states. The resolution of this spectrum is too coarse (250~\kms; see Table~\ref{tab:stisobs}) to separate the blended \ion{S}{3}/\ion{Si}{2} and \ion{S}{2}/\ion{Si}{2} lines; however, since the oscillator strengths of the \ion{Si}{2} lines in these blends are 1-2 orders of magnitude stronger than those of the S lines and we detect no \ion{S}{2} $\lambda$1251 or \ion{S}{2} $\lambda$1254 absorption in our spectrum, these blended lines primarily reflect the properties of the \ion{Si}{2} ions. Likewise, these values are the combined equivalent widths of all three velocity components detected in other metals in this sightline, as the resolution of this spectrum is too coarse to separate the components found by \citet{s91} and \citet{t99}, which have a maximum separation of $\Delta v = 160 \pm 10$~\kms. Therefore, we analyzed the higher resolution (20-25~\kms) G140M spectra to search for these velocity components. Measured radial velocities and equivalent widths for the species detected in the G140L spectrum are shown in Table~\ref{tab:stisparams}.

\begin{deluxetable}{lclcc}

\tablecolumns{5}
\tablewidth{0pt}

\tablecaption{Best-Fit Parameters to the Lines Detected in the STIS Spectra
\label{tab:stisparams}}

\tablehead{\colhead{Ion} & \colhead{$\lambda_{\rm rest}$} & \colhead{$cz$\tablenotemark{a}} & \colhead{$\log{N}$\tablenotemark{b}} & \colhead{W$_{\lambda}$} \\ & \colhead{(\AA)} & \colhead{(\kms)} & \colhead{(cm$^{-2}$)} & \colhead{(\AA)}}

\startdata
\ion{S}{3}/\ion{Si}{2} \dotfill & 1190.2, 1190.4 &
               $2230 \pm 100$ & \nodata        & $2.0 \pm 0.3$ \\
\ion{Si}{3} \dotfill & 1206.5 &
               $1510 \pm 10$  & $17.6 \pm 0.1$ & $1.7 \pm 0.3$ \\
\lya\ \dotfill & 1215.7 &
               $1430 \pm 20$  & $20.0 \pm 0.1$ & $6.5 \pm 2.4$ \\
\ion{S}{2}/\ion{Si}{2} \dotfill & 1259.5, 1260.4 &
               $1550 \pm 120$ & \nodata        & $1.2 \pm 0.2$ \\
\ion{O}{1} \dotfill & 1302.2 &
               $1520 \pm 110$ & \nodata        & $0.4 \pm 0.1$ \\
\ion{Si}{2} \dotfill & 1304.4 &
               $1350 \pm 210$ & \nodata        & $1.6 \pm 0.3$ \\
\ion{C}{2} \dotfill & 1334.5 &
               $1390 \pm 60$  & \nodata        & $1.4 \pm 0.1$ \\
\\
\siiv\ \dotfill & 1393.8, 1402.8 &
               $1375 \pm 20$  & $13.0 \pm 0.9$ & $0.1 \pm 0.1$ \\
            && $1435 \pm 15$  & $16.8 \pm 0.5$ & $0.4 \pm 0.2$ \\
            && $1535 \pm 10$  & $15.8 \pm 1.5$ & $0.4 \pm 0.2$ \\
\siiv\ Total:               &&& $16.8 \pm 0.8$ & $0.9 \pm 0.3$ \\
\\
\ion{Si}{2} \dotfill & 1526.7 &
               $1580 \pm 50$  & \nodata        & $1.3 \pm 0.4$ \\
\\
\civ\ \dotfill & 1548.2, 1550.8 &
               $1375 \pm 20$  & $16.2 \pm 2.2$ & $0.3 \pm 0.2$ \\
            && $1455 \pm 20$  & $18.2 \pm 0.5$ & $1.1 \pm 0.6$ \\
            && $1560 \pm 20$  & $15.9 \pm 2.3$ & $0.4 \pm 0.2$ \\
\civ\ Total:                &&& $18.2 \pm 0.8$ & $1.8 \pm 0.7$ \\
\\
\ion{Al}{2} \dotfill & 1670.8 &
               $1430 \pm 90$  & \nodata        & $1.2 \pm 0.3$ \\
\enddata

\tablenotetext{a}{\ Tabulated recession velocities are heliocentric.}
\tablenotetext{b}{\ Lines without a reported column density were measured from the G140L spectrum. All other lines were measured from the G140M spectra.}

\end{deluxetable}
 \clearpage

\begin{figure}[!t]
\begin{center}
\epsscale{0.95}
\plotone{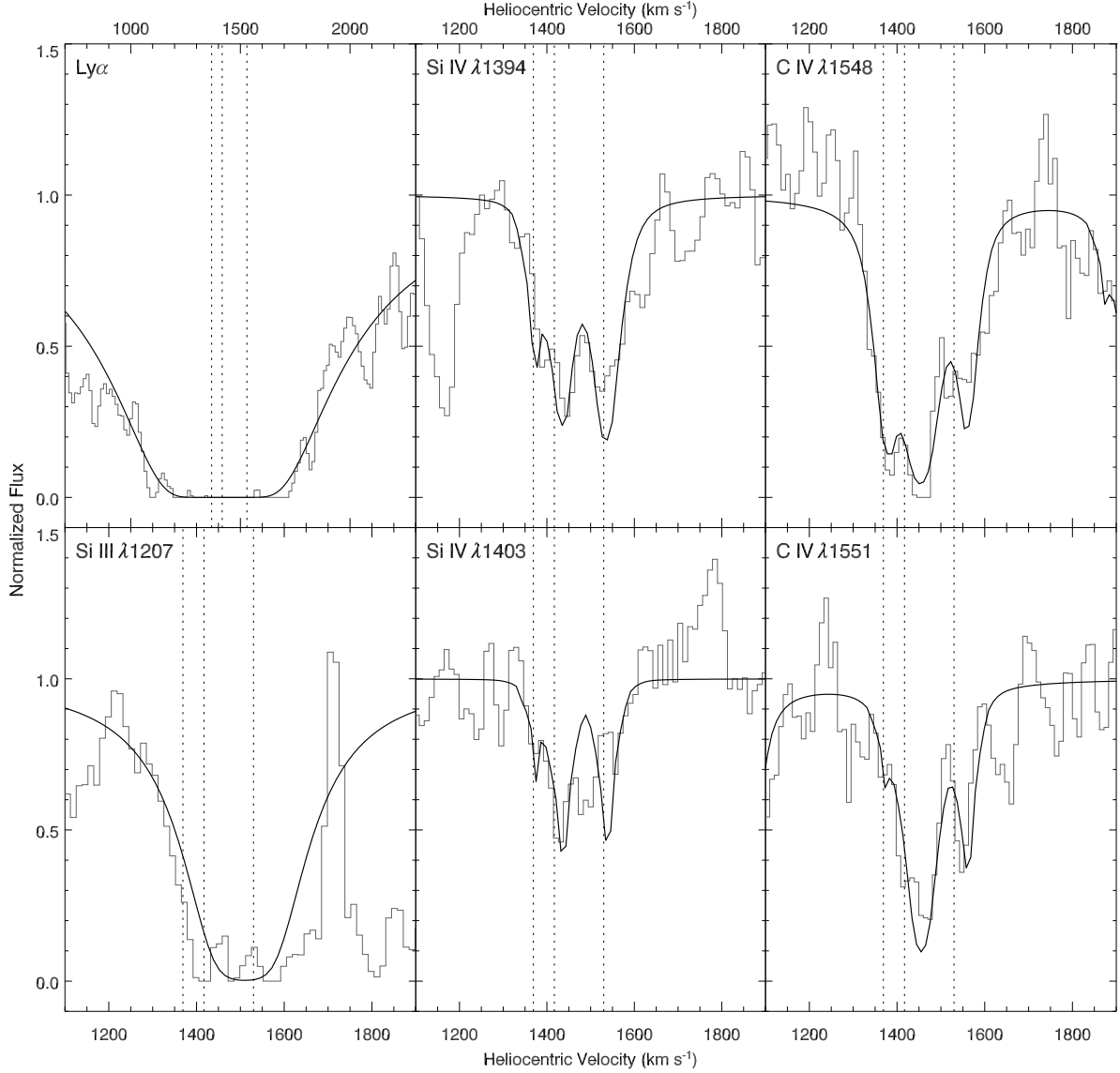}
\end{center}
\caption{Voigt profiles generated using the recession velocity and column density values from Table~\ref{tab:stisparams} and overlaid on the STIS G140M spectra. Each panel is labelled with the absorption feature that it depicts. The dotted vertical lines show the three velocity components at $cz = 1369$, 1417, and 1530~\kms\ found in \caii\ H \& K and \nai\ D by \citet{s91}. Note that the \lya\ panel shows a more extended velocity range than the other panels because the profile is black over nearly all of the narrower velocity range.
\label{fig:stisfits}}
\end{figure}

Voigt profiles were fitted to the \lya, \ion{Si}{3}, \siiv, and \civ\ absorption features in the G140M data. While the \lya\ absorption shows no sign of velocity structure, the small velocity offsets associated with the multiple metal-line components would not be detectable due to the wide damped \lya\ profile. No clear velocity substructure is present in the \ion{Si}{3} $\lambda$1207 absorption feature from \ngc\ at $\lambda \approx 1213$~\AA\ either, although the trough of the absorption feature spans velocities from $cz \approx 1400$-1650~\kms\ (see Figure~\ref{fig:stisfits}). The expected location of the \ion{N}{1} $\lambda$1200 absorption feature at the redshift of \ngc\ lies in the middle of the Galactic \ion{Si}{3} $\lambda$1207 absorption, so we cannot comment on its strength. 

In order to determine the most robust values of the recession velocity and the column density, the observed \lya\ profile was fit by a Voigt profile with constant $b$-value and varying $cz$ and $\log{N}$ in order to minimize the $\chi^2$ of the fit. The equivalent width of the best-fit profile was calculated from the corresponding model parameters, not from a direct non-parametric measurement of the absorption. The best-fit recession velocity, column density, and equivalent width for the \lya\ profile are shown in Table~\ref{tab:stisparams} and the corresponding Voigt profile is shown in the top-left panel of Figure~\ref{fig:stisfits}. The best-fit recession velocity of the \lya\ profile agrees with the \hi\ 21~cm velocity to within the errors. The best-fit column density ($N_{\rm HI} = 10^{20.0}$~cm$^{-2}$) makes this absorber a Lyman limit system, which are typically associated with the halos of bright galaxies \citep{steidel95,steidel98,stocke05}. Together with the \hi\ 21~cm radio data, the \lya\ column density yields estimates of the spin temperature from the single dish and VLBA data of $435 \pm 140$~K and $590 \pm 380$~K, respectively. The \ion{Si}{3} profile was fit in the same way as the \lya\ profile and its best-fit recession velocity, column density, and equivalent width are shown in Table~\ref{tab:stisparams}. The corresponding Voigt profile is shown in the bottom-left panel of Figure~\ref{fig:stisfits}.

In contrast to the \hi\ 21~cm and \lya\ data, the \civ\ and \siiv\ profiles have multiple velocity components as seen in several other metal species in the \qso\ sightline \citep{s91,t99}. Multiple Voigt profiles were fit to the \siiv\ and \civ\ doublets in order to determine the most robust values of recession velocity and column density. The initial velocity estimates for these profiles were $cz = 1370$, 1420, and 1530~\kms\ \citep[chosen to match the velocity components at $cz = 1369 \pm 2, 1417 \pm 2$, and $1530 \pm 10$~\kms\ found in \nai\ and \caii\ by][]{s91}.  A range of $b$-values from 4-20~\kms\ was chosen and held constant while the recession velocity and column density were allowed to vary in order to minimize the $\chi^2$ of the fits to the \siiv\ and \civ\ profiles. The best-fit parameters from the fits with different $b$-values were then compared to estimate the fitting errors. The resulting best-fit recession velocities, column densities, and equivalent widths are shown in Table~\ref{tab:stisparams}. The corresponding Voigt profiles for the \siiv\ and \civ\ doublets are shown in the middle and right panels of Figure~\ref{fig:stisfits}, respectively. The best-fit velocities for the 1370 and 1530~\kms\ components in both the \siiv\ and \civ\ profiles agree with the values found by \citet{s91} to within the combined errors. The best-fit velocity at $cz = 1435 \pm 15$~\kms\ in the \siiv\ profile is just barely discrepant from the \citet{s91} value of $cz = 1417 \pm 2$~\kms\ at the $1\sigma$ level, while the best-fit velocity at $cz = 1455 \pm 20$ in the \civ\ profile differs from the \citet{s91} velocity by $1.8\sigma$. Considering that the \siiv\ and \civ\ components are significantly blended \citep[although much less so than \ion{Mg}{2};][]{t99}, all of the best-fit velocities agree well with the component velocities determined previously for \caii\ and \nai\ \citep{s91}.

\section{Discussion}
\label{discussion}

Perhaps the most remarkable aspect of this sightline is the consistent velocity structure from neutral elements (\nai\ D, \ion{Mg}{1}) to singly-ionized species (\caii\ H \& K, \ion{Mg}{2}, \ion{Fe}{2}) to high ions (\siiv, \civ). \citet{s91} found three velocity components in the profiles of \nai\ D and \caii\ H \& K and determined that these components lie at heliocentric recession velocities of $cz = 1369 \pm 2$, $1417 \pm 2$, and $1530 \pm 10$~\kms. \citet{t99} used HST/GHRS to examine the absorption-line profiles in the wavelength region $\lambda = 2250$-3000 \AA\ and found absorption lines of \ion{Mg}{1}, \ion{Mg}{2}, and \ion{Fe}{2} associated with \ngc. The \ion{Mg}{1} $\lambda$2853 profile has velocity components at heliocentric velocities of $cz = 1370$ and 1420~\kms\ and the \ion{Fe}{2} $\lambda$2344, $\lambda$2374, and $\lambda$2600 profiles all show broad absorption centered at $cz = 1450$~\kms\ and extending from $cz \approx 1350$-1600~\kms, although no clear velocity substructure is present (the \ion{Fe}{2} $\lambda$2586 line was also observed, but due to blending with Galactic \ion{Fe}{2} $\lambda$2600, its velocity structure is indeterminate). Like the \lya\ profile described above, the \ion{Mg}{2} $\lambda\lambda$2796, 2800 profiles found by \citet{t99} are highly saturated. Thus, while absorption is present in these profiles from $cz=1300$-1600~\kms, it is impossible to detect, or rule out, any velocity substructure in these transitions. Finally, the present study uses HST/STIS spectra to examine the \siiv\ and \civ\ absorption lines associated with \ngc\ and finds evidence of multiple velocity components in these profiles. We find three distinct velocity components in both doublets, and by fitting Voigt profiles to the observed absorption we have found that the velocity components for the \siiv\ and \civ\ transitions generally match the components found by \citet{s91} to within the combined errors. We believe, but cannot prove definitively due to line blending, that the velocities of \siiv\ and \civ\ match the low ion velocities exactly.

Because all of the far-UV absorption in Table~\ref{tab:stisparams} is in the \lya\ forest region for this QSO ($z = 0.533$), line blending with low-$z$ \lya\ interlopers (e.g., the line just blueward of \siiv\ $\lambda$1394 in Figure~\ref{fig:normspec}) can contaminate the wavelength region of interest and create false velocity components. \lya\ interlopers are especially problematic for the \siiv\ and \civ\ profiles due to the sensitivity of the STIS G140M data. Nonetheless, the presence of a consistent velocity structure over numerous ionization states implies that the high ionization gas is spatially coincident with the neutral and singly-ionized gas; i.e., the gas exists in three ``clouds'' above the disk of \ngc. 

The Na/Ca ratio for these absorbers \citep[$N$(\nai)$/N$(\caii$) \approx 1$;][]{s91} is not consistent with gas originating in the Galactic disk \citep[$N$(\nai)$/N$(\caii$) \approx 2$-200;][]{crawford89}, but no studies have measured the Na/Ca ratio out to a Galactocentric radius of 30\hinv~kpc, the galactocentric radius of the \ngc\ clouds if they lie in the disk of \ngc\ rather than its halo. However, \citet{morton86} found a tentative cutoff in the distance at which \caii\ could be detected from a galaxy center of $\sim {\rm D}(0)_{25}$, the isophotal face-on diameter of a galaxy at a surface brightness of 25~mag~arcsec$^{-2}$ (${\rm D}(0)_{25} = 12$\hinv~kpc for \ngc) and \citet{briggs80} found that disk gas detected in \hi\ 21~cm at galactocentric radii $> 1.5$ Holmberg radii is uncommon in the local Universe and that the distribution of the disk gas shows a sharp cutoff at $N_{\rm HI} < 10^{19}~{\rm cm}^{-2}$. These results suggest that the \ngc\ clouds are likely associated with the halo (i.e., not the very extended disk) of the galaxy. We propose that each of the velocity components in the \qso\ sightline is caused by an absorbing cloud of gas above the plane of \ngc\ that is analogous to a Galactic HVC. In \S\ref{disc:gaskin} we compare the kinematics of the absorbing clouds around \ngc\ and Galactic HVCs, in \S\ref{disc:spintemp} we compare their spin temperatures and cloud sizes and in \S\ref{disc:balance} we discuss the ionization conditions necessary to produce the observed equivalent width ratios.

\begin{figure}[!t]
\begin{center}
\epsscale{0.95}
\plotone{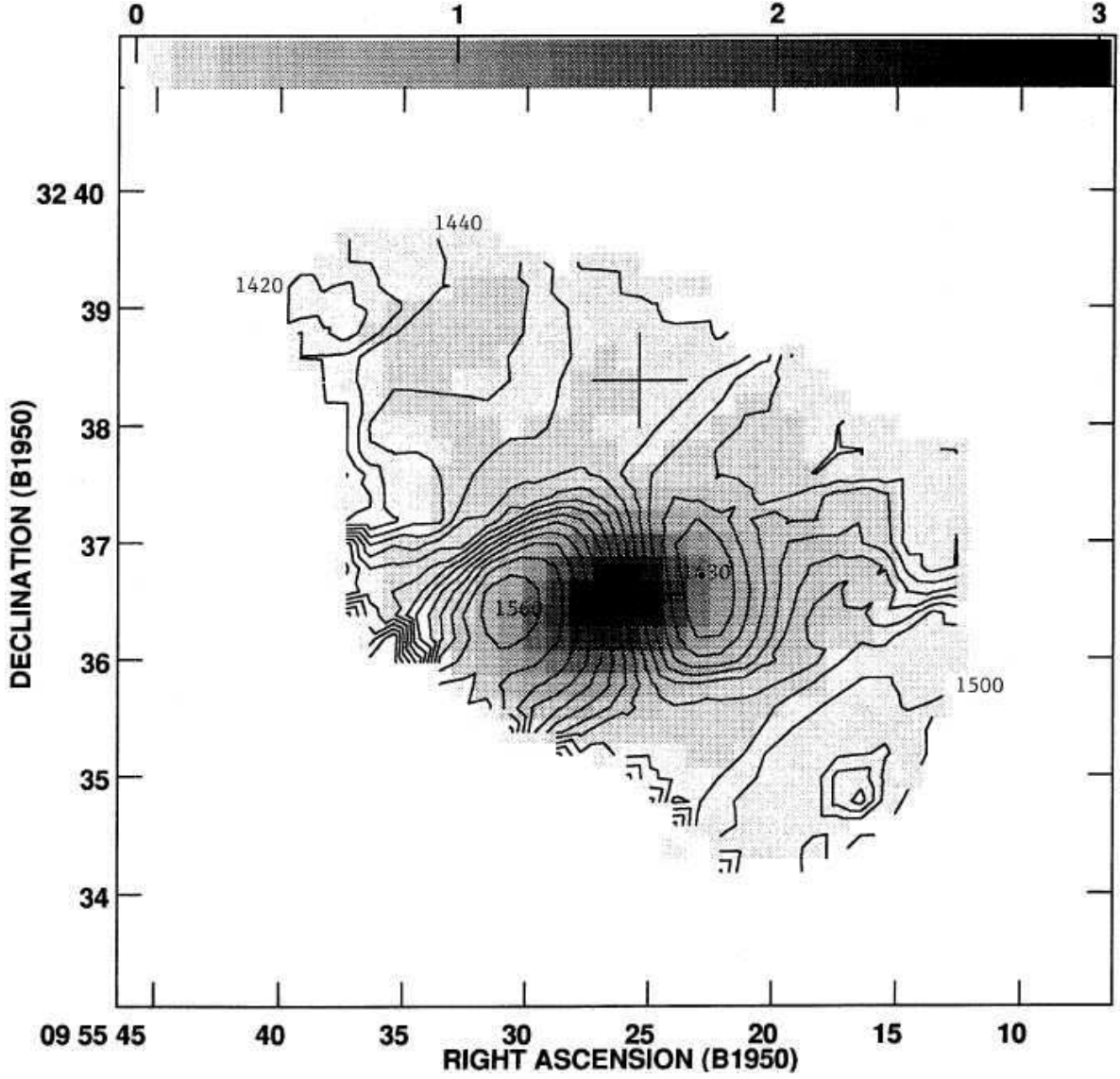}
\end{center}
\caption{\hi\ emission map and rotation curve of \ngc\ reproduced from \citet{cvg}. The gray scale shows the integrated \hi\ column density for \ngc\ at $50\arcsec$ (5\hinv~kpc) resolution, ranging from $2.2 \times 10^{19}$~atoms~cm$^{-2}$ ({\it white}) to $1.3 \times 10^{21}$~atoms~cm$^{-2}$ ({\it black}). The contours show the intensity-weighted mean velocity for the \hi\ gas. Contour levels are 1430, 1440, ..., 1560~\kms. A cross marks the position of \qso.
\label{fig:cvg}}
\end{figure}

\subsection{Gas Kinematics}
\label{disc:gaskin}

The \hi\ 21~cm emission map and rotation curve of \citet{cvg} (their Fig.~7) is reproduced in Figure~\ref{fig:cvg}, where the halo gas to the NE (SW) can be seen to have systematically lower (higher) radial velocities than the disk ``below'' these regions. If the halo gas in these regions is rotating similarly to disk gas, then the northern half is closer to us than the southern half (i.e., we are looking at the ``underside'' of the galaxy). Using the systemic velocity for \ngc\ of $1465 \pm 5$~\kms\ \citep{cvg}, an inclination of $68^{+4}_{-3}$~degrees \citep{rubin82}, and assuming that the clouds are on purely radial trajectories along the minor axis of \ngc\ as one might expect for an outflowing starburst wind, the 1370, 1420, and 1530~\kms\ clouds are moving at $-260^{+50}_{-70}$, $-130^{+30}_{-50}$, and $+170^{+90}_{-50}$~\kms\ with respect to the nucleus of \ngc, respectively. Thus, given the galaxy's orientation, two of the clouds are infalling and one is outflowing. The uncertainties in the above velocities take into account the uncertainty in the velocity of each component and the uncertainty in the systemic velocity and inclination of \ngc, but they do not take into account uncertainties in our physical model. If the clouds are moving on radial orbits that are not along the minor axis of \ngc\ then their inflow/outflow velocities could be much larger than this if they are moving nearly along the plane of the sky. Of course, these clouds could also share some or all of the rotation of the disk, in which case the radial inflow/outflow velocity is diminished.

The rotation curve in Figure~\ref{fig:cvg} yields a dynamical mass for \ngc\ of $1.2 (\pm 0.2) \times 10^{11}$\hinv~M$_{\sun}$, which leads to an escape velocity of $310 \pm 30$~\kms\ at a radius of 11~kpc. This velocity implies that even if our assumed geometry is wrong and the halo gas is counter-rotating with respect to the disk (i.e., the southern side of \ngc\ is closer to us than the northern side) the \hi\ detected cloud at $cz = 1420$~\kms\ would still be bound to \ngc\ if it is traveling on a radial orbit along the galaxy's minor axis. Especially considering that these clouds are likely to share some rotational velocity with the underlying disk, it is likely that all three clouds are bound to the galaxy irrespective of the assumed geometry, lending more weight to the supposition that these clouds are not the result of an outflowing galactic wind. Despite the modest starburst (1.4~M$_{\Sun}$~yr$^{-1}$) in \ngc, there is no strong evidence that halo gas is escaping, and since these velocities are shared by both the low and high ions, this conclusion is not based solely upon the velocity of the neutral halo gas. Note in Figure~\ref{fig:cvg} that the \qso\ sightline samples a chord close to the minor axis of \ngc, the expected location of an outflowing starburst wind; e.g., \citet*{heckman90} find that the high velocity gas in several nearby starbursts can be explained by a bi-conic outflow with a typical opening angle of $\sim 65\degr$. The location of \qso\ relative to \ngc\ ($\approx 10$-15$\degr$ from the minor axis; see Figure~\ref{fig:cvg}) places this sightline well within this cone angle.

The usual convention for Galactic HVCs is to call gas with a radial velocity $\ga 90$~\kms\ with respect to the local standard of rest (LSR) an HVC; however, the range of radial velocities allowed by differential galactic rotation varies with Galactic position \citep{wakker97}. Converting the LSR velocity to the Galactic standard of rest (GSR), an HVC with $v_{LSR} = 90$~\kms\ would have $v_{GSR}$ in the range $-130$ to $+310$~\kms\ \citep{wakker97}. However, it should be stressed that this $v_{GSR}$ is a radial velocity only and does not represent the full space velocity of the HVC. The observed radial $v_{GSR}$ for the 1370, 1420, and 1530~\kms\ clouds ($-95$ to $+65$~\kms) are clearly in this velocity range, and so have kinematics and locations consistent with Galactic HVCs. Additionally, many Galactic HVCs \citep[e.g., Complex C;][]{tripp03} are observed to be infalling onto our local Galactic disk. Since \ngc\ also shows evidence for infalling clouds and its starburst is of comparable strength to those of other bright (${\rm L} \sim {\rm L}^*$) local starburst galaxies, one would expect that their superwinds might not be able to escape their gravitational potentials either.

We have measured a column density of $\log{N_{\rm HI}} = 20.0 \pm 0.1$~cm$^{-2}$ for the 1417~\kms\ cloud, and have calculated an upper limit of $N_{\rm HI} < 9 \times 10^{18}$~cm$^{-2}$ for the 1370 and 1530~\kms\ clouds, assuming that they have a similar spin temperature and velocity width as the 1420~\kms\ cloud. Large angle sky surveys such as \citet{bajaja85} and \citet{hulsbosch88} search for Galactic HVCs down to a detection limit of $N_{\rm HI} \approx 2$-$3 \times 10^{18}$~cm$^{-2}$, so it is questionable whether the 1370 and 1530~\kms\ clouds would be detected in such a survey if they were located near the Milky Way. The combined \citet{bajaja85} and \citet{hulsbosch88} surveys find a covering factor for Galactic HVCs, as seen from Earth, of $\sim 0.4$ \citep{wakker91}. Thus, if \ngc\ has the same covering factor of HVCs as the Milky Way, there is a $\sim 64\%$ probability of intercepting at least one cloud with $N_{\rm HI} \ga 2 \times 10^{18}~{\rm cm}^{-2}$ along the sightline to \qso, assuming that the clouds are randomly distributed. Thus the coincidence of one strong HVC and two weak ones in this sightline is consistent with the expectations from our Galaxy's HVC population.

Recently, \citet{thilker04} have discovered an extensive population of \hi\ clouds surrounding M~31 with the Green Bank Telescope. They find $\ga 20$ discrete clouds within 50~kpc of the M~31 disk that they believe are analogs to Galactic HVCs. The M~31 clouds have radial velocities comparable to the velocity of the outer disk of M~31 and masses of $(0.15$-$1.3) \times 10^6$~M$_{\Sun}$ \citep{thilker04}. Additionally, \citeauthor{thilker04} find a filamentary halo component concentrated near the M~31 systemic velocity that extends at least 30~kpc from its nucleus. If the extensive \hi\ cloud systems of the Milky Way and M~31 are ubiquitous in the low-$z$ Universe then it is no surprise that we have detected similar clouds around \ngc.

\subsection{Spin Temperature and Cloud Size}
\label{disc:spintemp}

Our dataset allows us to make several estimates of the spin temperature of the \hi\ in the clouds. From our VLBA observations (\S\ref{obs:vlba}) we have calculated a column density for \hi\ absorption toward \qso\ of $N_{\rm HI}/T_{\rm s} = 1.7 (\pm 0.4) \times 10^{17}$~cm$^{-2}$~K$^{-1}$. Combining this result with the column density of \hi\ emission ($8(\pm 4) \times 10^{19}$~cm$^{-2}$) found by \citet{cvg} (see Figure~\ref{fig:cvg}), we calculate a spin temperature of $T_{\rm s} = 470 \pm 260$~K for the strongest \hi\ component seen on the line of sight to \qso. By fitting a Voigt profile to the \lya\ absorption associated with \ngc\ in our HST/STIS spectrum of \qso\ (\S\ref{obs:stis}), we have found a column density of $\log{N_{\rm HI}} = 20.0 \pm 0.1$~cm$^{-2}$ for the 1420~\kms\ cloud, yielding a spin temperature of $T_{\rm s} = 590 \pm 380$~K when combined with the VLBA $N_{\rm HI}/T_{\rm s}$ measurement. A third value is obtained by comparing the single-dish Arecibo optical depth with the \lya\ column density: $435 \pm 140$~K. While this latter value has the smallest statistical errors, the radio and UV beam sizes against which the absorption is measured are not as good a match (radio continuum size $\ga 30$~mas; UV continuum size $< 1$~mas). In contrast, the VLBA $T_{\rm s}$ value uses only the core of \qso, which has a size of $\approx 10$~mas. However, within the rather large measurement errors, all of these values agree.

Unfortunately, very little is known about the spin temperatures of Galactic HVCs, but \citet{akeson99} have measured a $3\sigma$ lower limit of $T_{\rm s} > 180$-320~K on the line of sight toward one Galactic HVC. The spin temperature measurements of the strongest absorbing cloud near \ngc\ are clearly consistent with this lower limit. On the other hand, \hi\ clouds in the Outer Arm of the Galactic disk (${\rm radius} \sim 15$-20~kpc) have spin temperatures that are much lower than those calculated for the 1420~\kms\ cloud \citep[typically $\sim 50$-150~K, c.f.,][]{akeson99}, further indicating that this cloud is not associated with the disk of \ngc.

We can also estimate the kinetic temperature ($T_{\rm k}$) of the \hi\ using the width of the 21~cm absorption profile. This profile has a FWHM of $4.2 \pm 0.1$~\kms\ in the Arecibo data, which corresponds to a $b$-value of $2.5 \pm 0.1$~\kms\ and yields $T_{\rm k} = 380 \pm 30$~K. If one performs the same calculation starting with the FWHM of the VLBA 21~cm absorption profile then $T_{\rm k} = 475 \pm 35$~K. These estimates agree with the spin temperature measurements above to within the errors, suggesting that $T_{\rm s}$ and $T_{\rm k}$ are coupled. To our knowledge, this extragalactic cloud is the first for which both $T_{\rm s}$ and $T_{\rm k}$ have been measured and found to be approximately equal. Particle collisions naturally drive the spin temperature of a gas toward the kinetic temperature. Using the collisional de-excitation rate coefficients of \citet{allison69}, a particle density of $n_{\rm HI} \ga 0.1~{\rm cm}^{-3}$ is required for collisions to be the dominant source of temperature coupling at this kinetic temperature \citep{kulkarni88}. \citet{t99} found a density of $n_{\rm HI} \approx 0.005~{\rm cm}^{-3}$ for the \ngc\ clouds from photoionization modeling, smaller than the density required for particle collisions to couple $T_{\rm s}$ and $T_{\rm k}$. 

Resonant scattering of \lya\ photons can also drive the spin temperature to the kinetic temperature, since the radiation field exchanges energy with the gas during the scattering process through atomic recoil \citep{field58,field59}. This interaction causes $T_{\rm s}$ to approach the color temperature of the radiation, which is equal to $T_{\rm k}$ if enough scatterings have occurred \citep[i.e., the optical depth is large enough;][]{field59,deguchi85}. \citet{deguchi85} found that \lya\ scattering dominates the excitation of low-density extragalactic clouds if the \lya\ optical depth ($\tau_L$) is $\ga 10^5$. Using Equation~7 of \citet{wolfe79}, $\tau_L$ in the \ngc\ clouds is $2.9(\pm 0.1) \times 10^7$. Since the optical depth is so large, the \lya\ photons cannot penetrate easily into the cloud from an external source; thus, the \lya\ photons must be generated inside the cloud itself from the recombination of electrons and protons to form \hi\ \citep{deguchi85,kulkarni88}. In statistical equilibrium, the recombination rate is equal to the ionization rate of \hi\ by cosmic rays and EUV and soft X-ray photons that are capable of penetrating to the cloud's interior. For resonant \lya\ scattering to drive $T_{\rm s}$ to within 10\% of $T_{\rm k}$ at this kinetic temperature, an ionization rate of $\ga 5 \times 10^{-17}$ ionizations per hydrogen atom per second is required \citep{deguchi85}. \citet{watson84b} found an ionization rate of $2.5 \times 10^{-18}$ ionizations per hydrogen atom per second for penetrating radiation from the extragalactic 1-10~keV X-ray flux. The true ionization rate in the \ngc\ clouds is likely sufficient to couple $T_{\rm s}$ and $T_{\rm k}$ (i.e., $\ga 5 \times 10^{-17}~{\rm s}^{-1}$) when one includes the contributions of cosmic rays, soft X-rays produced in \ngc, the extragalactic X-ray background, and some Lyman continuum photons escaping from \ngc.

The diversity of our observational data also allows us to estimate the size of the strongest absorbing cloud. Since absorption from the 1420~\kms\ cloud is present to comparable optical depths in the spectra of both the core and jet regions of \qso\ (see Figure~\ref{fig:HIspec}), the absorbing cloud must be larger than the separation between the two regions, 20~mas (see Table~\ref{tab:vlbafits}). At the redshift of \ngc\ this angular distance corresponds to a physical distance of 2\hinv~pc.  Since our estimates of $T_{\rm s}$ using either the \lya\ $N_{\rm HI}$ value or the \hi\ 21~cm emission $N_{\rm HI}$ value \citep{cvg} agree to within errors, there is little evidence for clumping on angular scales smaller than the VLA beam size of $50\arcsec$. Otherwise, one would expect a significant discrepancy between the $T_{\rm s}$ calculated from the large VLA beam and the STIS pencil beam due to the filling factor of the gas. Finally, a visual examination of Figure~\ref{fig:cvg} reveals no noticeable clumping in \hi\ emission or \hi\ velocity on angular scales slightly larger than $50\arcsec$. Thus, a reasonable estimate of the angular size of the absorbing clouds is $\sim 50\arcsec$, which corresponds to a physical size of 5\hinv~kpc at the distance of \ngc. This characteristic size yields a mean physical density for this cloud of $\sim 0.007~{\rm cm}^{-3}$, remarkably similar to the $n_{\rm HI} = 0.004$-0.006~cm$^{-3}$ found from photoionization modeling by \citet{t99}. This agreement is further evidence that the absorbing cloud is rather uniform on a 5\hinv~kpc scale.

It is surprising that a spherical cloud with a 5\hinv~kpc radius would have a line width that is purely thermal with no contribution from turbulent or bulk motions. Galactic HVCs tend to have larger line widths than that found in \hi\ 21~cm absorption for the 1420~\kms\ cloud; the $b$-value for a typical Complex C sightline is $\approx 10$~\kms\ for low ions like \hi\, \ion{O}{1}, \ion{Fe}{2}, and \ion{S}{2} \citep[see][]{collins03,sembach04}. It is unclear why the \hi\ 21~cm line width in the 1420~\kms\ cloud is so narrow ($b = 2.5 \pm 0.1$~\kms). Perhaps the cloud is sheet-like rather than spherical such that it is smaller along the line of sight than in the transverse direction.

Based upon new H$\alpha$ emission detections, \citet{putman03} find that most Galactic HVCs lie within $\sim 40$~kpc of the Galaxy and are not extragalactic members of the Local Group. The Galactic HVC Complex C spans $\sim 90\degr$ on the sky and lies at a distance of 2-14~kpc \citep[depending on the spiral arm with which one assumes the HVC is associated;][]{putman03}, yielding a physical size of 3-22~kpc. This result implies that the absorbing clouds around \ngc\ not only lie at comparable galactocentric distances as Galactic HVCs but also have comparable physical sizes.

\subsection{Ionization Balance}
\label{disc:balance}

\citet{t99} found that all of the low and intermediate ionization species in the \qso\ sightline could be explained by a photoionization model with an extragalactic ionizing flux of $\Phi_{\rm ion} \geq 2600$~photons~cm$^{-2}$~s$^{-1}$ assuming a power-law spectrum with index 1.8 and no ionizing flux from \ngc\ \citep[if one allows a significant contribution from \ngc\ then it becomes impossible to constrain the extragalactic radiation field;][]{t99}. \citet{t99} also placed an upper limit on the escape fraction of ionizing photons from \ngc\ ($f_{esc} \le 0.02$) based on H$\alpha$ images of the \qso/\ngc\ system. This value is comparable to the escape fraction for Galactic HVCs ($f_{esc} = 0.01$-0.02) measured by \citet{putman03}. The photoionization model of \citet{t99}, which fits all of the low ions (\ion{Na}{1} through \ion{Mg}{2} and \ion{Fe}{2}) predicts \civ\ and \siiv\ equivalent widths of $\sim 10$~m\AA, which would be undetectable in the G140M spectra presented here. Thus, the easily detected higher ions are much more likely to be collisionally-ionized by shocks or some other process despite sharing the same kinematics with the lower ions. 

In addition, the photoionization modeling of \citet{t99} sets a modest lower limit on metallicity for the ensemble of all three of these clouds of $[Z/Z_{\Sun}] \ge -0.6$ ($Z \ge 0.25$\Zsun). This limit is specifically derived from the \ion{Si}{2} 1304 \AA\ line (see Table~\ref{tab:stisparams}) but is consistent with the observed strengths of other detected lines. This limit was derived assuming the extreme case of a purely extragalactic ionizing spectrum with an ionizing flux of ${\rm I}_{\nu} = 1 \times 10^{-23}~{\rm ergs~s^{-1}~cm^{-2}~Hz^{-1}~sr^{-1}}$ as inferred by \citet{t99}; however, if we assume the opposite extreme of an ionizing spectrum that comes purely from ionizing photons escaping the plane \ngc\ with $f_{esc} = 0.02$, the metallicity limit does not change appreciably. The metallicity limit for the \ngc\ clouds of $Z \ge 0.25$\Zsun\ is comparable to recently measured metallicities of Galactic HVCs. \citet{tripp03} used a STIS echelle spectrum of the background quasar 3C~351 to find a metallicity of $Z = 0.1$-0.3\Zsun\ for Complex C and \citet{sembach02} used a {\it Far Ultraviolet Spectroscopic Explorer} (FUSE) spectrum of the background quasar Ton~S120 to find a 3$\sigma$ metallicity limit of (O/H$) < 0.46$~solar for the compact high velocity cloud CHVC~224.0$-$83.4$-$197. \citet*{collins04} find similar metallicities of $Z = 0.34 \pm 0.14$\Zsun\ and $Z = 0.06^{+0.12}_{-0.04}$\Zsun\ for the highly ionized HVCs detected toward PKS~2155$-$304 at velocities of $v_{LSR} = -140$ and $-270$~\kms, respectively.

Highly ionized gas has been detected in several Galactic HVCs. \citet{fox04} have detected Complex C in a similar range of ionic species to those we detect for the clouds toward \qso. Using spectra from HST/STIS and FUSE, they found that \ion{O}{6}, \ion{Si}{2}, and \hi\ are all coincident in velocity, mimicking the ionization structure that we see in the \ngc\ clouds. By modeling the high ion column density ratios $N$(\siiv)/$N$(\ion{O}{6}), $N$(\civ)/$N$(\ion{O}{6}), $N$(\ion{N}{5})/$N$(\ion{O}{6}), \citeauthor{fox04} conclude that neither a single collisional ionization equilibrium nor photoionization equilibrium model can produce the highly ionized gas observed in Complex C. They hypothesize that the highly ionized gas is a collisionally ionized skin that forms in the conductive or turbulent interfaces that arise when the predominantly photoionized cloud moves through a surrounding hot medium. Given the similar ionization structure between Complex C and the \ngc\ clouds, it is plausible that the same phenomena produce the ions in both; i.e., the low ions observed by \citet{s91} and the intermediate ions observed by \citet{t99} have strengths and ratios consistent with a single photoionization model but not the \siiv\ and \civ, which require another ionization source. While \citeauthor{fox04} discuss several alternative hypotheses for the observed ionization structure, the analogy between Complex C and these clouds is retained. We note that the inferred velocities of the three HVC-like clouds in \ngc\ relative to a static halo in that galaxy yield shock temperatures easily sufficient to produce large amounts of \civ\ and \siiv.

\citet*{lehner01} have detected the Galactic HVC 291.2$-$41.2+80 along the line of sight to the Magellanic Bridge star DI 1388 in several species that we have detected toward \qso, including \ion{C}{2}, \ion{O}{1}, \ion{Si}{2}, \ion{Si}{3}, \siiv, and \ion{S}{3}. However HVC 291.2$-$41.2+80 has not been detected in \hi\ to a 5$\sigma$ limit of $10^{19}$~cm$^{-2}$ \citep{lehner01}, which suggests that this Galactic HVC is analogous to the 1370 and 1530~\kms\ clouds around \ngc. Using the \ion{C}{2}/\ion{Si}{2} and \ion{O}{1} column densities to trace the hydrogen column of HVC 291.2$-$41.2+80 yields estimated column densities of $\log{(N_{\rm HI}+N_{\rm HII})} = 17.62~{\rm cm}^{-2}$ and $\log{N_{\rm HI}} = 16.24~{\rm cm}^{-2}$, respectively, which \citeauthor{lehner01} argue indicates that much of the hydrogen is ionized since \ion{C}{2} and \ion{Si}{2} trace both neutral and singly-ionized gas whereas \ion{O}{1} traces only neutral gas. These Galactic analogs suggest that the 1370 and 1530~\kms\ clouds could also be significantly ionized and not predominantly neutral as is the case for HVCs with higher \hi\ column densities.

The highly ionized HVCs detected by \citet{sembach99} toward the QSOs Mrk 509 and PKS 2155$-$304 have a different ionization structure than the \ngc\ clouds or HVC 291.2$-$41.2+80 in that they are detected exclusively in high ions such as \civ, \siiv, and \ion{N}{5} with no detectable absorption from low or intermediate ions (e.g., \ion{Si}{2} or \ion{C}{2}) or \hi\ 21~cm emission at corresponding velocities. A search for 21~cm emission near these sightlines revealed \hi\ HVCs at comparable velocities to the highly ionized clouds within $\sim 2\degr$ of the sightlines \citep[the \hi\ HVC with the highest column density has $N_{\rm HI} < 2 \times 10^{18}~{\rm cm}^{-2}$;][]{sembach99}, which suggests that the highly ionized gas traces the extended ionized regions of the \hi\ HVCs in much the same way as the highly ionized skin proposed by \citet{fox04} for Complex C. \citeauthor{sembach99} also find that the lack of intermediate ions in their highly ionized HVCs argues against collisional ionization as the primary ionization mechanism for these clouds; instead they prefer a photoionization model in which the source possesses a hard QSO-like spectrum as the principle photoionization mechanism, although they don't rule out a contribution to the ionization balance from collisionally ionized gas. The requirement of a QSO-like source in their photoionization models implies that the highly ionized HVCs of \citeauthor{sembach99} lie at large distances from the Galactic plane \citep[30-200~kpc if the highly ionized HVCs and their \hi\ counterparts are in thermal pressure equilibrium;][]{sembach99} where they are exposed to the extragalactic radiation field. 

Subsequent observations of these HVCs with FUSE \citep{sembach03} and STIS \citep{shull03,kraemer03} have changed the initial interpretation of the \citet{sembach99} GHRS data. Strong \ion{O}{6} absorption was found in the FUSE spectra of these HVCs at column densities that are too large for reasonable photionization models to accomodate. Many intermediate ions are detected in the STIS spectra, including \ion{Si}{2}, \ion{Si}{3}, \ion{C}{2}, and \ion{C}{3}; \hi\ was even detected in two HVCs toward PKS~2155$-$304 at $N_{\rm HI} = 10^{15}$ to $10^{16}~{\rm cm}^{-2}$ \citep{collins04}. \citet{collins04} argue that the \ion{O}{6} in these HVCs is collisionally ionized and the presence of absorption in ions ranging from \hi\ to \civ, \siiv, and \ion{O}{6} suggests that these HVCs are multiphase. They propose that the high ions in these clouds arise from hot gas that is potentially produced behind bow shocks from clumps falling toward the Galactic plane. This interpretation is remarkably similar to that of \citet{fox04} for Complex C and suggests a similar source for the high ions detected in the \ngc\ clouds.

\section{Conclusions}
\label{conclusions}

We have presented sensitive new radio and ultraviolet spectra to determine if the multiple velocity components ($cz = 1317 \pm 2, 1417 \pm 2, 1530 \pm 10$~\kms) found in \nai\ and \caii\ by \citet{s91} and \ion{Mg}{1}, \ion{Mg}{2}, and \ion{Fe}{2} by \citet{t99} toward \qso\ are also present in \hi\ \lya, \siiv, and \civ. Our \hi\ 21~cm data from the VLBA and Arecibo show evidence for only the strongest (1420~\kms) velocity component and set upper limits, $N_{\rm HI} < 9 \times 10^{18}~{\rm cm}^{-2}$, on the column density of the other two components. The 1420~\kms\ absorber is a Lyman limit system with  a column density of $N_{\rm HI} = 10^{20.0}~{\rm cm}^{-2}$ based on Voigt profile fitting of the \lya\ absorption feature in our STIS G140M spectra. Lyman limit systems are generally associated with the halos of luminous galaxies \citep{steidel95,steidel98,stocke05}, and indeed our best value of the measured spin temperature for this absorber ($T_{\rm s} = 435 \pm 140$~K) is significantly larger than the spin temperatures \citep[$\sim 50$~K;][]{akeson99} found in the Galactic disk and comparable to the lower limit of $T_{\rm s} > 180$-320~K found toward one Galactic high velocity cloud \citep[HVC;][]{akeson99}. The Na/Ca ratio for these absorbers is not consistent with gas in the Galactic disk either \citep{s91}, further suggesting that these absorbers are located in the halo, not the extended disk, of \ngc.

While the 1420~\kms\ component is still the only component detected in \hi, we did detect all three velocity components in both lines of the \siiv\ and \civ\ doublets in our STIS G140M spectra. The presence of all three velocity components in various metal species and ionization states implies that the highly ionized gas is spatially coincident with the neutral and singly ionized gas. We hypothesized that the three velocity components correspond to three clouds of gas $\sim 11$\hinv~kpc from the nucleus of \ngc\ that are analogous to Galactic HVCs. We have found several lines of evidence that support this argument:
\begin{itemize}
\item We find that galactocentric radial velocities of the \ngc\ clouds ($-95$ to +65~\kms) lie in the velocity range expected for a Galactic HVC with $v_{LSR} = 90$~\kms. 
\item Assuming a cloud location along the minor axis of \ngc\ and purely radial motion, the most likely kinematic solution for the $10^{20}~{\rm cm}^{-2}$ cloud and the underlying disk of \ngc\ require this cloud to be falling toward the disk as frequently observed for Galactic HVCs.
\item Our estimated cloud size of 5\hinv~kpc is comparable to the estimated size of the Galactic HVC Complex C.
\item The spin temperature of the \ngc\ clouds ($T_{\rm s} = 435 \pm 140$~K) is consistent with the lower limits found for Galactic HVCs \citep[$T_{\rm s} \ga 200$~K;][]{akeson99}.
\item The photoionization modeling of \citet{t99} places a lower limit on the metallicity of the \ngc\ cloud ensemble of $Z \ge 0.25$\Zsun, which is similar to the metallicities derived along several sightlines toward Galactic HVCs.
\item We detect the same intermediate ions in the \ngc\ clouds (e.g., \ion{C}{2}, \ion{Si}{2}) that are detected in many highly ionized Galactic HVCs.
\item The upper limit on the escape fraction of UV radiation to these clouds is comparable to that found from the Galaxy to  Galactic HVCs \citep{putman03}. A much larger escape fraction would be expected along the minor axis of a starburst wind \citep*[see][]{dove00,maclow99}. 
\end{itemize}
Thus, the overall physical picture of the \qso/\ngc\ system is that of a Galactic ``fountain'', not an outflowing starburst wind.

We find no evidence that halo gas along this sightline is escaping the galaxy's gravitational potential, despite the fact that \ngc\ is a modest starburst with a star formation rate of 1.4~M$_{\Sun}$~yr$^{-1}$. The \qso\ sightline is located near the minor axis of \ngc, which is where one would expect to find signatures of an outflowing starburst wind \citep[e.g., ][]{heckman90}. Nevertheless, regardless of whether we assume that the halo gas shares the rotation direction of the disk, the deprojected velocities of the \ngc\ clouds are all less than the escape velocity from the gravitational potential well if they are travelling on purely radial orbits along the minor axis of \ngc. Furthermore, the most consistent analysis of the disk/halo gas kinematics from a detailed \hi\ emission map \citep{cvg} determines that two of these clouds (including the \hi\ 21~cm detected cloud) are infalling. For either kinematic solution (halo gas rotating or counter-rotating) the $N_{\rm HI} = 10^{20}~{\rm cm}^{-2}$ system is gravitationally bound to \ngc. Thus, if these clouds truly are analogous to Galactic HVCs, one would expect that Galactic HVCs are also bound to the Galactic gravitational potential rather than that of the Local Group and that Galactic HVCs are located closer to the Galaxy than the Local Group barycenter.

If instead we take the perspective that at least one of the velocity components observed in \ngc\ towards \qso\ is an outflowing starburst wind, do we observe what might be expected for such a circumstance? Almost certainly not. First, we might expect the higher ions to have different kinematics from the low ions; the latter would be expected to be seen at lower galactocentric velocities as cool clumps in a hotter, more highly ionized wind. While the kinematic solution does strictly allow an outflowing, even escaping wind if the geometry is just right (e.g., ejection nearly along the plane of the sky), the more straightforward kinematic solution is that two of the clouds are infalling, certainly not expected in a starburst wind scenario. The limit on the escape fraction of ionizing radiation from \ngc\ ($f_{esc} \le 2\%$) is also much lower than expected in a situation where an outflowing starburst wind would be expected to clear away almost all of the neutral material above the disk \citep{dove00,maclow99}. The large inferred size of 5\hinv~kpc for the neutral $N_{\rm HI} = 10^{20}~{\rm cm}^{-2}$ cloud is far larger than might be expected for a cool clump in a starburst wind; i.e., this size is nearly half of the cone opening angle ($\sim 65\degr$) suggested by \citet{heckman90}. Thus, a starburst model is a poor match to many of the observations for these clouds described in this paper. 

This result also suggests that the superwinds produced by other comparably massive local starbursts (e.g., NGC~253, NGC~4945) might not be capable of escaping their galaxy's potential well either. However, starburst winds may escape the shallower gravitational potential wells of less massive galaxies \citep[${\rm L} < 0.1$\Lstar;][]{martin03,stocke04}. Even though all of the low, intermediate, and high ions detected in these putative HVCs share the same kinematics, there remains the possibility that even more highly ionized gas could possess different kinematics that trace a wind escaping from \ngc. Unfortunately, at F$_{\lambda}(1200$~\AA$) \approx 4$-$5 \times 10^{-15}~{\rm ergs~s^{-1}~cm^{-2}}$~\AA$^{-1}$, \qso\ is too faint in the far-UV to be observed easily with FUSE, which could be used to test this hypothesis by observing the \ion{O}{6} $\lambda\lambda$1032, 1038 absorption lines at high signal to noise ratio.

The QSO/galaxy pair PKS~1327-206/ESO~1327-2041 shares many characteristics with the \qso/\ngc\ system. PKS~1327-206 and ESO~1327-2041 have a projected separation of 13\hinv~kpc \citep{cvg}. ESO~1327-2041 has a similar luminosity to \ngc\ \citep[${\rm L} \approx 1.2$\Lstar;][]{lauberts89,marzke94} and was noted by \citet{giraud86} and \citet{cvg} to possess a polar ring of extraplanar gas. Thus, it is worthwhile to re-examine the gas kinematics and morphology of this QSO/galaxy pair in light of the HVC interpretation presented here for the similar \qso/\ngc\ system.

The results we present here strongly support the model for Galactic HVCs that places them within a few kpc of the disk of the Milky Way \citep[e.g.,][]{bland-hawthorn02,putman03} and not at $\sim 1$~Mpc as ``extragalactic clouds'' within the Local Group \citep{blitz99}. In the case of \ngc\ there is little doubt that these absorbers are within a few kpc of its galactic plane and have remarkably similar properties to those of Galactic HVCs. Deep \hi\ 21~cm searches for HVC analogs in nearby galaxy groups \citep{verheijen01,zwaan01} and in non-targeted \hi\ surveys \citep{zwaan00} have failed to find extragalactic \hi\ clouds, which would be expected if the \citet{blitz99} hypothesis were correct. The present result adds to the significant evidence against the extragalactic evidence for HVCs. However, \citet{cvg} note that the very extended \hi\ in \ngc\ and the possibly complex kinematics of the extra-planar \hi\ are quite similar to the \hi\ morphologies seen in interacting galaxy systems (e.g., M~51 \& Mrk 348), although \ngc\ has no obvious close companion. Could it be that the Milky Way and M~31 are also unusual in possessing such an extensive system of extra-planar \hi\ clouds?

{\acknowledgments The authors thank T. Ghosh and C. Salter for their help with the Arecibo observations and data reduction, T. Rector for his help with the VLBA reductions, and P. Maloney for helpful discussions. B. A. K. acknowledges support from NASA Graduate Student Researchers Program grant NGT5-154. B. A. K. and J. T. S. acknowledge support from NASA HST General Observer grants GO-09506 and GO-06593. J. T. is supported by the Department of Astronomy and Astrophysics at the University of Chicago.

\end{document}